\documentclass[aps,prd,reprint,nofootinbib,showkeys,showpacs,superscriptaddress,longbibliography,floatfix]{revtex4-1}

\pdfoutput=1

\usepackage{amsmath}
\usepackage{amssymb}
\usepackage[american]{babel}
\usepackage{booktabs}
\usepackage{dcolumn}  
\usepackage[T1]{fontenc}  
\usepackage{graphicx}  
\usepackage[colorlinks]{hyperref}  
\usepackage[utf8]{inputenc}  
\usepackage{multirow}
\usepackage{txfonts}  
\usepackage{url}
\usepackage[dvipsnames]{xcolor}
\usepackage{float}
\usepackage{lipsum}
\newcolumntype{d}[1]{D{.}{.}{#1}}
\newcolumntype{v}[1]{D{,}{,\ }{#1}}


\renewcommand {\arraystretch}{1.5}

\hypersetup{
	citecolor = blue,
	linkcolor = RoyalPurple,
	urlcolor = blue
}

\begin{document}

	\title{Bayesian Comparison of the Cosmic Duality Scenarios}
	
	\author{W. J. C. da Silva}
	\email{williamjouse@fisica.ufrn.br}
	\affiliation{Departamento de Astronomia, Observatório Nacional, Rio de Janeiro - Rio de Janeiro, 20921-400, Brasil}
	\affiliation{Universidade Federal do Rio Grande do Norte, Departamento de F\'{\i}sica, Natal - Rio Grande do Norte, 59072-970, Brasil}

	\author{R. F. L. Holanda}
	\email{holandarfl@fisica.ufrn.br}
	\affiliation{Universidade Federal do Rio Grande do Norte,
		Departamento de F\'{\i}sica, Natal - Rio Grande do Norte, 59072-970, Brasil}

	\author{R. Silva}
	\email{raimundosilva@fisica.ufrn.br}
	\affiliation{Universidade Federal do Rio Grande do Norte,
		Departamento de F\'{\i}sica, Natal - RN, 59072-970, Brasil}
	\affiliation{Universidade do Estado do Rio Grande do Norte, Departamento de F\'{\i}sica, Mossor\'o - Rio Grande do Norte, 59610-210, Brasil}

	
	\date{\today}

	\begin{abstract}
		The cosmic distance duality relation (CDDR), $D_{\rm L}(1+z)^{-2}/D_{\rm A}=\eta=1$, with $D_{\rm L}$ and $D_{\rm A}$, being the luminosity and angular diameter distances, respectively, is a crucial premise in cosmological scenarios. Many investigations try to test CDDR through observational approaches, even some of these ones also consider a deformed CDDR, i.e., $\eta=\eta(z)$.  In this paper, we use type Ia supernovae luminosity distances and galaxy cluster measurements (their angular diameter distances and gas mass fractions) in order to perform a Bayesian model comparison between $ \eta(z) $ functions. We show that the data here used are unable to pinpoint, with a high degree of Bayesian evidence, which $\eta(z)$ function best captures the evolution of CDDR.
	\end{abstract}

	\maketitle
	
	\section{Introduction}\label{sec:intro}

	\begin{figure*}[t]
		\centering
		\includegraphics[scale=0.363]{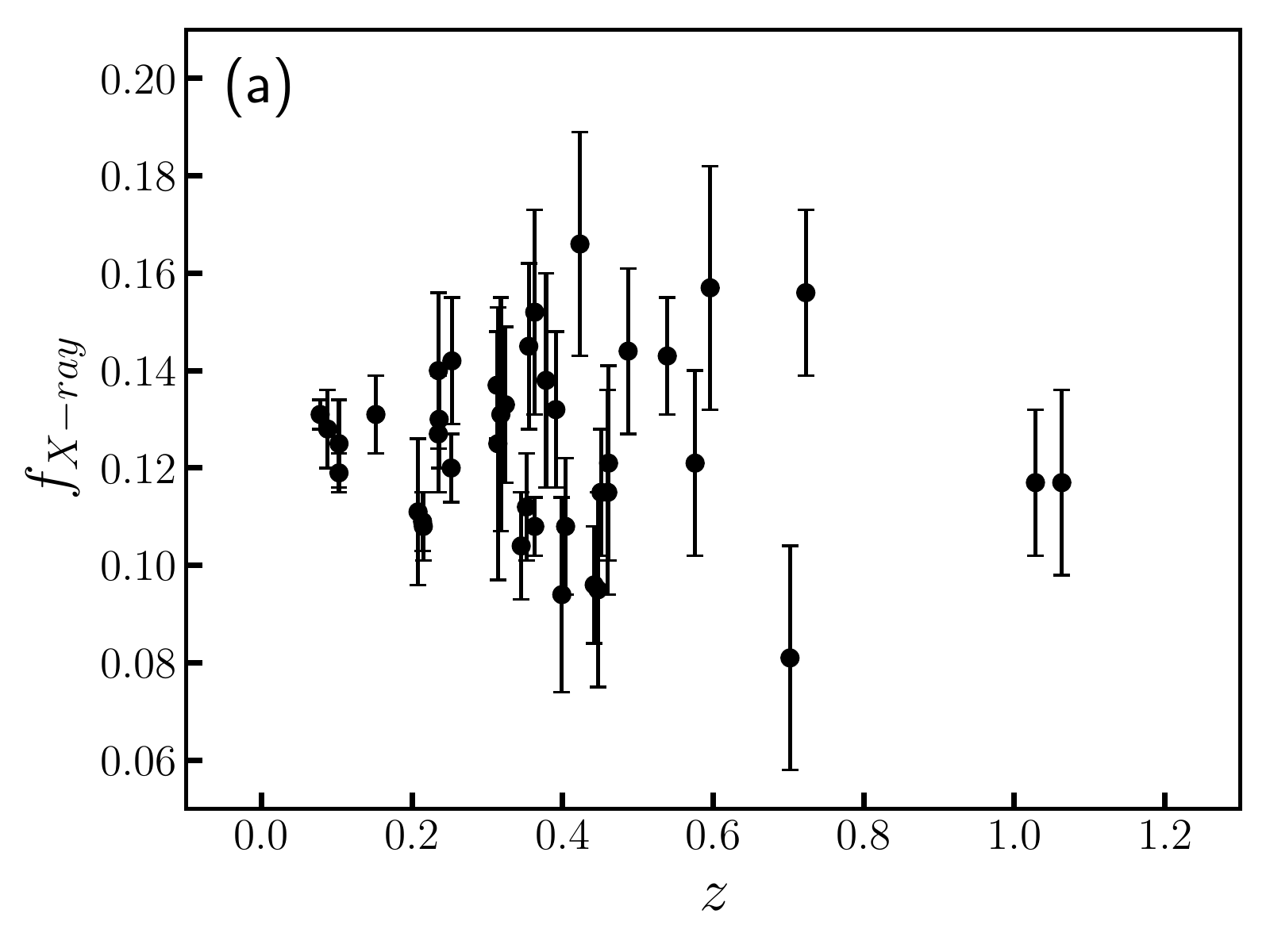}
		\includegraphics[scale=0.363]{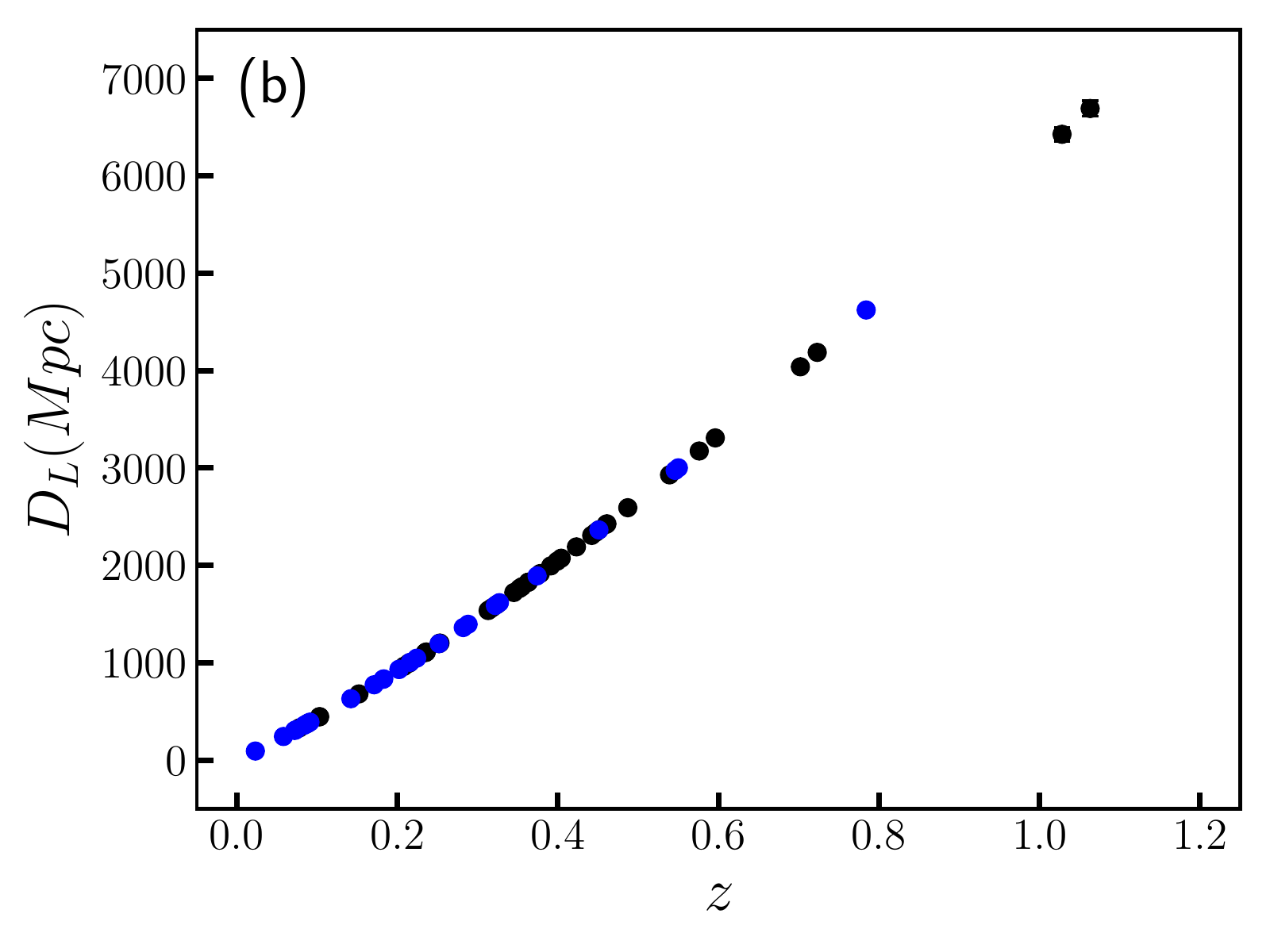}
		\includegraphics[scale=0.363]{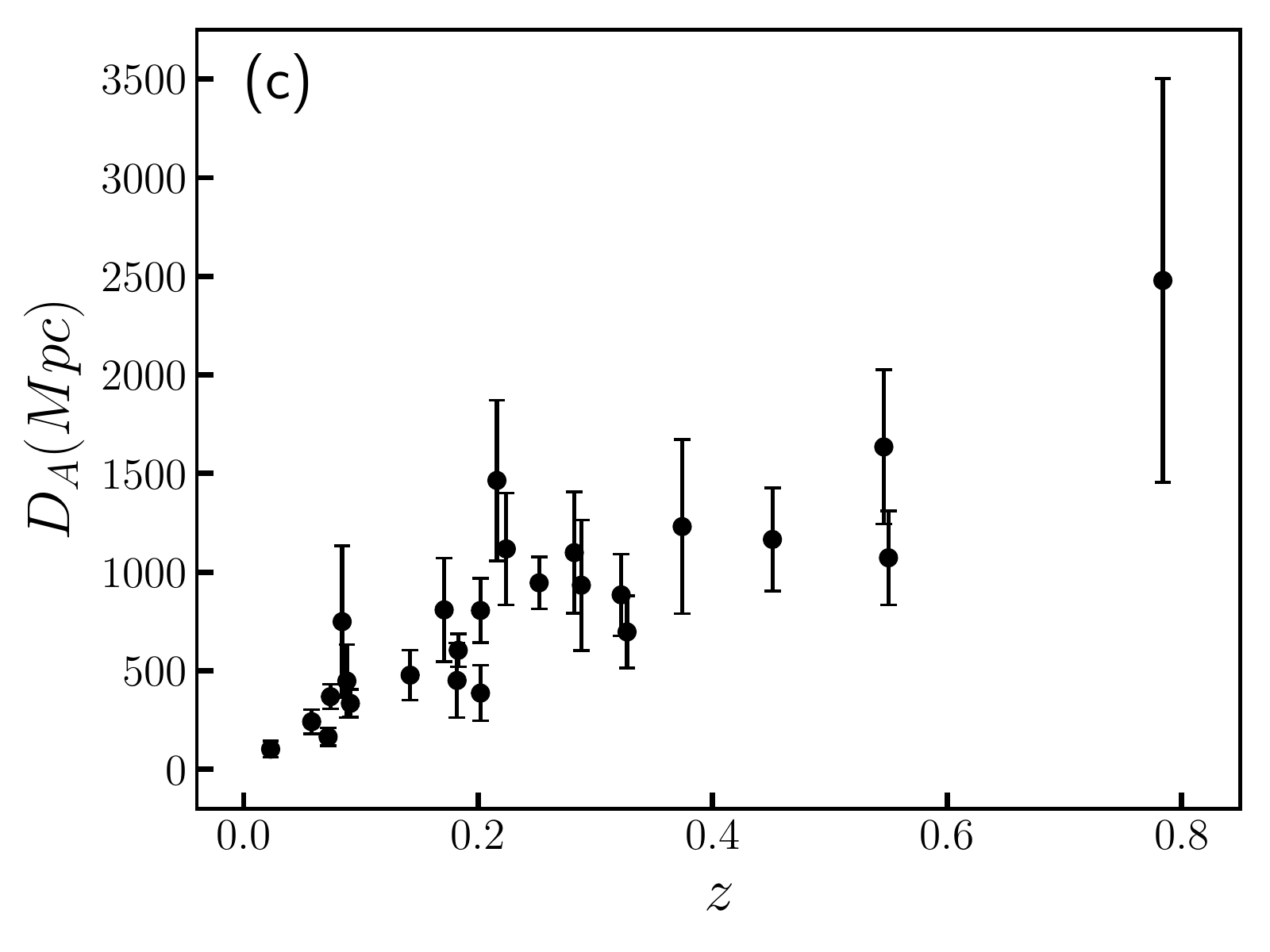}
		\caption{\label{fig:data-fig} The panels show the data used in this work. Part (a) shows $f_{\text{X-ray}}$, the Part (b) shows the SNe Ia data measure at the same redshift of the x-ray mass fraction (black) and, the angular diameter distances of the galaxy clusters (blue). Part (c) shows the angular distance of the galaxy clusters.}
	\end{figure*}
	
	Measuring distances in cosmology is of crucial importance when one wants to relate observational data with theoretical models. The two types of distance most used in cosmology are the luminosity distance, $D_{\rm L}$, and the angular diameter distance, $D_{\rm A}$. The former is a distance measurement associated with an object based on the decrease of its brightness, and the latter one is associated with the measurement of the angular size of the object projected on the celestial sphere. These cosmological distances are functions of the redshift $z$ of the astronomical object considered and are connected by a relation known as the cosmic distance duality relation (CDDR), $\frac{D_{\rm L}(z)}{D_{\rm A}(z)(1+z)^{2}} = \eta= 1$, or as Etherington's reciprocity law in the background of the astronomical observation \cite{Etherington1933,Etherington2007}. 
	
	The CDDR is obtained in the context of Friedmann-Lemaître-Roberton-Walker metric but holds for general metric theories of gravity in any background, in which photons travel in null geodesics and, the number of photons is conserved during cosmic evolution \cite{Bassett2004}. In fact, the generality this relationship is the crucial importance in the context of observational cosmology. { Briefly, if the gravity is a  metric theory and  if the Maxwell equations are valid  the distance duality relation is satisfied ($\eta=1$) \cite{Ellis2009}. Then, a little deviation from $\eta=1$ may indicate the possibility of a new physics, such as, photon coupling with particles beyond the standard model of particle physics\footnote{For instance, axion-photon conversion induced by intergalactic magnetic fields \cite{Avgoustidis2010,Jaeckel2010}.}, variation of fundamental constants, and scalar fields with a nonminimal multiplicative coupling to the electromagnetic Lagrangian, among others \cite{Bassett2004,Avgoustidis2010,Jaeckel2010,Hees2014}.  The presence of systematic errors in observations, such  as photon absorption by dust, also can violate the CDDR validity \cite{Bassett2004,Avgoustidis2010}.}
	
	Simultaneously with the increase in the number and the quality of astronomical data, different methods have been proposed to test the validity of the CDDR.  We can divide them in two classes: cosmological model-dependent tests based on the $\Lambda$ cold dark matter ($\Lambda$CDM) framework \cite{DEBERNARDIS2006,Uzan2004,Avgoustidis2010,Holanda2011,More:2016fca,Piazza2016} and cosmological model-independent ones. The last ones have been performed by using combinations of several astronomical data: angular diameter distance of galaxy clusters, galaxy cluster gas mass fraction\footnote{In the Ref. \cite{Holanda2016b} only massive galaxy clusters observations were 
    considered in its method to test the CDDR.}, type Ia supernovae (SNe Ia), strong gravitational lensing, cosmic microwave background, gamma ray bursts, radio compact sources, \textit{cosmic microwave background radiation}, baryon acoustic oscillations, gravitational waves, etc. \cite{Holanda2010,Lima2011,Li2011,Gonalves2011,Meng2012,Holanda2012,Yang2013,Liang2013,Shafieloo2013,Zhang:2014eux,SantosdaCosta2015,Jhingan2014,Chen2015,Holanda2016,Rana2016,Liao2016,Holanda2016b,Holanda2017,Rana2017,Lin2018,Fu2019,Ruan2018,Holanda2019,Chen2020,Zheng2020,Kumar:2020ole,Hu2018}.
	
	In order to test the CDDR, the basic approach has been to consider a deformed expression, given by $\frac{D_{\rm L}(z)}{D_{\rm A}(z)(1+z)^{2}} = \eta(z)$, and to obtain constraints on some $\eta(z)$ functions.\footnote{The Ref. \cite{SantosdaCosta2015}, by applying a non-parametric method, namely, Gaussian process, proposed a test based on galaxy clusters observations and $H(z)$ measurements (see also the Ref. \cite{Chen2015}) without using $\eta(z)$ functions.} In this context, the authors of the  Ref. \cite{Holanda2010}  assumed two $\eta(z)$  functions, such as: $\eta(z) = 1 + \eta_0z$ and  $\eta(z) = 1 + \eta_0z/(1+z)$. Actually, these functions are clearly inspired by similar expressions for the equation of state parameter of
    dark energy models. By using angular diameter distance samples of galaxy clusters jointly with luminosity distances of SNe Ia, they obtained  that CDDR is valid within $2\sigma$ ($\eta_0\approx 0$). However, other $\eta(z)$ functions were also proposed, e.g., $\eta(z) = \eta_0 + \eta_1z$, $\eta(z) = \eta_0 + \eta_1z/(1+z)$, $\eta(z) = \eta_0 + \eta_1 \ln(1+z)$,
	and $\eta(z)=(1+z)^{\epsilon}$ (see, for instance, Refs. \cite{Avgoustidis2010,Liang2013,Li2011}). As a basic result from literature, the CDDR validity has been verified, at least, within $2\sigma$ c.l.. However, it is very worth it to stress that the current analyses cannot distinguish which $\eta(z)$ function better describes the data.
	
	Recently, the Bayesian inference has been widely used as an useful tool in order to search several problems in physics \cite{vonToussaint2011}, cosmology, and astronomy \cite{Hobson2002,Trotta2007,Liddle2007,Trotta2008,Santos2017,Lonappan2018,Kerscher2019,Silva2019,daSilva2019,Cid2019,Cuceu:2020dnl}. Specifically, the Bayesian inference has been a powerful technique to study many issues in cosmology, e.g., effects of bulk viscosity in dark matter and dark energy \cite{Silva2019,daSilva2019}, by comparing the dark
    energy models \cite{Santos2017,Lonappan2018} and and interacting in the dark sector \cite{Cid2019} . This technique has been implemented to analyze other
    cosmological models (see Refs. \cite{Kilbinger2010,Leclercq2018}).	An interesting question here is try to analyze which $\eta(z)$ should be viable from the Bayesian inference standpoint.
	
	In order to face this issue, in this paper, we use SNe Ia luminosity distances and galaxy cluster measurements (angular diameter distances and gas mass fractions) to perform a Bayesian model comparison between $ \eta(z) $ functions used in literature, such as: $\eta=\eta_0$, $\eta(z)=1+ \eta_1 z$, $\eta(z)=1+ \eta_1z/(1+z)$, $\eta(z)=\eta_0 + \eta_1z$ and  $\eta(z)=\eta_0 + \eta_1z/(1+z)$. The basic idea is to estimate the Bayesian evidence and compute the Bayes factor of each $\eta(z)$ function with respect to $\eta=\eta_0$. The $\eta$ constant is chosen as the standard model because if $\eta_0=1$ the standard CDDR is recovered and we obtain this value within $2\sigma$ C.L. with the dataset used in our analyses. 
	
	This paper is divided in the following way. In Sec. \ref{sec:data}, we describe the data used in this work: SNe Ia and galaxy clusters observations. In Sec. \ref{sec:methodology}, we present the equations that describe the cosmological data. In Sec. \ref{sec:parameters}, we show the $\eta(z)$ parametrizations assumed in this work. Next, in Sec. \ref{sec:beyesian}, we achieve the Bayesian analysis by considering the data and parametrizations previously presented. Finally, the Sec. \ref{sec:results} presents the main results of the statistical analysis, and Sec. \ref{sec:conclusions}, we show the conclusions of the work.

	\section{Data}\label{sec:data}
	
	\subsection{Type Ia supernovae}\label{subsec:sneia}
	
	The luminosity distances are obtained  from the SNe Ia sample called Pantheon \cite{Scolnic2018}. The full compilation consists of $1049$ spectroscopically confirmed SNe Ia and covers a redshift range of $0.01 \leq z \leq 2.3$, being  the most recent wide refined sample of SNe Ia. However, to perform the appropriate tests on the CDDR, SNe Ia at the same (or approximately) redshift of the galaxy clusters must be used (see below). Then, for each galaxy cluster, we make a selection of SNe Ia according to the criterion: $|z_{GC}-z_{\text{SNe Ia}}| \leq 0.005$. Then, we perform the weighted average by for each galaxy cluster by:
	
	\begin{equation}
	\bar{\mu} = \frac{\sum_i \mu_i/\sigma_{\mu_i}^{2}}{\sum_i 1/\sigma_{\mu_i}^{2}},
	\end{equation}
	
	\begin{equation}
	\sigma_{\bar{\mu_i}}^2 = \frac{1}{\sum_i 1/ \sigma_{\mu_i}^{2}},
	\end{equation}
	where $\mu_i (z)$ is the distance module of SNe Ia. Hence, the luminosity distance follows from $D_{\rm L} (z) = 10^{(\bar{\mu}-25)/5}$ $[Mpc]$, and its error is given by error propagation, $\sigma_{D_{\rm L}}^{2} = (\partial D_{\rm L} / \partial \bar{\mu} )^2\sigma_{\bar{\mu}}^{2}$ [see Fig. \ref{fig:data-fig}(b)].

\subsection{Galaxy clusters}\label{subsec:gas}
	
In order to perform the analyses, we also use two different observations of galaxy clusters, namely: angular diameter distance  and gas mass fraction. The dataset is:
\begin{itemize}
\item The $D_{\rm A}(z)$ data of $25$ galaxy clusters obtained via their Sunyaev-Zeldovich effect (SZE) {\it plus} x-ray observations and presented by the Ref. \cite{DeFilippis2005}. The sample is distributed over the redshift interval $0.023 \leq z \leq 0.784$ [see Fig. \ref{fig:data-fig}(c)]. As it is largely known, it is possible to take advantage of the different cluster electronic density dependencies in these phenomena and with some assumptions about  morphology of galaxy cluster evaluate its angular diameter distance, given by \cite{Cavaliere1978,Bonamente2006,Reese2002,DeFilippis2005}
\begin{equation}
D_{\rm A}(z)\propto \frac{(\Delta T_{0})^{2}\Lambda_{eH0}}{S_{X0}T_{e}^{2}}\frac{(1+z)^{-4}}{\theta_{c}},
\end{equation}
where $\theta_{c}$ is its  core radius obtained from  SZE and x-ray analysis, $S_{X0}$ is  central X-ray surface brightness, $\Delta T_{0}$ is the central SZE, $T_e$ is the electronic temperature of cluster,  $\Lambda_{eH0}$ is the cooling factor, and $z$ is  the redshift of the galaxy cluster.

It is important to stress that various sources of uncertainty  (statistical and  systematic) are present in this technique, such as isothermality, morphology of the cluster, x-ray and SZE calibrations, presence of clumps,  kinetic SZE, radio halos, cosmic microwave background (CMB) anisotropy, and X-ray background. These contributions of errors added in quadrature give an error of $\approx 20$\% (statistical) and $\approx 12-15$\% (systematic) on the angular diameter distance estimated (see details in Refs. \cite{Bonamente2006,Reese2002}). We have added in quadrature the statistical and systematic errors.  Relativistic corrections also need to be taken into account for hot galaxy clusters\cite{Itoh1998}. Since the spherical assumption to describe galaxy clusters has been severely questioned for the Chandra satellite observations, the x-ray surface brightness of the clusters considered here was described by the elliptical $\beta$-model \cite{DeFilippis2005}.

Besides, it is important to comment that the SZE is redshift independent only if there is no injection of photons into the CMB. For this case, the CMB temperature evolution law is given by $T_{\rm CMB}(z) = T_0(1 + z)$.  On the other hand, if there is a departure of the CDDR validity, the CMB  temperature evolution law is modified, and the SZE becomes redshift dependent \cite{Battistelli2002,Luzzi2009}; consequently, the angular diameter distances estimated for the clusters need to be corrected. However, the SZE observations considered in Ref. \cite{DeFilippis2005} were performed in $30$ GHz; in this band, the effect on the SZE from a variation of $T_{\rm CMB}(z)$ is completely negligible \cite{Melchiorri2005} (see Ref. \cite{Holanda2017b} for the case where the SZE observations are performed in $150$ GHz).

\item The most recent x-ray mass fraction measurements of $40$ galaxy clusters in redshift range $0.078 \leq z \leq 1.063$ from Ref. \cite{Mantz2014}, Fig. \ref{fig:data-fig}(c). These authors measured the gas mass fraction in spherical shells at radii near $r_{2500}$, rather than integrated at all radii (less than $r_{2500}$) as in previous works. As consequence,  the theoretical uncertainty in the gas depletion obtained from hydrodynamic simulations is reduced \cite{Mantz2014,Planelles2013} [see Fig. \ref{fig:data-fig}(a)].
\end{itemize}

	\section{Methodology}\label{sec:methodology}
	
	In this section, we present the equations used in our analyses. It is important to stress that previous works discussed how the expression $D_{\rm L}(1+z)^{-2}/D_{\rm A}=\eta=1$ has to be modified if one wishes to test it by using x-ray and SZE observations of galaxy clusters \cite{Holanda2016b}. These observations are affected if there are deviations from the CDDR validity and variation in the fine structure constant. In the following, we discussed briefly this point (see details in Refs. \cite{Hees2014,Holanda2016b,Holanda2016c,Holanda2017}).
	
	\subsection{$D_{\rm A}$ from galaxy clusters and SNe Ia}
	
	In Ref. \cite{Hees2014} was studied as modifications of gravity by the presence of a scalar field with a coupling electromagnetic Lagrangian affect cosmological signatures, e.g., CDDR validity and variations of the fine structure constant. In this context, Ref. \cite{Holanda2016b} discussed how the angular diameter distance of a galaxy cluster obtained from its SZE and x-ray observations ($D_{\rm A}^{\text{data}}$) is affected by a such multiplicative coupling. The authors showed that the observations  do not give the true distance, but $D_{\rm A}^{\text{data}}=\eta^4(z)D_{\rm A}$.  Then, if one wants to test the CDDR  by using $D_{\rm L}(1+z)^{-2}D_{\rm A}^{-1}=\eta$ and galaxy clusters data, the angular diameter distance $D_{\rm A}(z)$ must be replaced by $D_{\rm A}(z)=\eta^{-4}D_{\rm A}^{\text{data}}$ (see Refs. \cite{Holanda2016b,Uzan2004} for more details). Then, the equation basic to test the CDDR using $D_{\rm L}$ from SNe Ia and $D_{\rm A}^{\text{data}}$ from galaxy clusters is
	
	\begin{equation}
	\frac{D_{\rm L}(z)}{(1+z)^2D_{\rm A}^{\text{data}}(z)}=\eta^{-3}(z),
	\end{equation}
	or, equivalently,
	\begin{equation}\label{eq:DA-DL}
	\eta_{\rm obs}(z)=\left(\frac{D_{\rm L}(z)}{(1+z)^2D_{\rm A}^{\text{data}}(z)}\right)^{-1/3}.
	\end{equation}

	\subsection{Gas mass fractions $\times$ SNe Ia}
	
	The test by using gas mass fraction performed here  is completely based on the equations obtained in Refs. \cite{Holanda2016c,Holanda2017}.\footnote{In Ref.\cite{Holanda2018} the effects from a possible cosmic opacity on the cosmological constrains obtained from gas mass fraction also were discussed.} Likewise, the authors of those references showed that the usual expression used in gas mass fraction measurements (where $\eta=1$; see Ref. \cite{Allen2007}) have to be replaced by 
	
	\begin{eqnarray}\label{GasFrac4}
	f^{\rm obs}_{\rm X-ray}(z) &=& N \left[\frac{\eta(z)^{7/2}D_{\rm L}^{*3/2}}{D_{\rm L}^{3/2}}\right],
	\end{eqnarray}
	if one wishes taking into account possible deviation of the CDDR validity and a variation of the fine structure constant. One may define yet
	
	\begin{eqnarray}\label{GasFrac5}
	\eta_{\rm obs}(z) &=& \left[\frac{f_{\rm gas}^{2/3}D_{\rm L}(z)}{N^{2/3}D_{\rm L}^{\rm *}}\right]^{3/7},
	\end{eqnarray}
	where the symbol * denotes quantities from a fiducial cosmological model used in the observations (usually a flat $\Lambda$CDM model where $\eta=1$). The $N$ factor corresponds to the parameters:  $K(z)$, which quantifies inaccuracies in instrument calibration, as well as any bias in the masses measured due to substructure, bulk motions and/or nonthermal pressure in the cluster gas, and $\gamma$, the depletion factor, which corresponds to  the ratio by which $f_{\rm gas}$ is depleted with respect to the universal baryonic mean and the $\Omega_{\rm b}/\Omega_{\rm M}$ ratio. The $K(z)$ parameter for this sample was estimated to be  $K=0.96 \pm 0.12$ (statistical $+$ systematic errors), and no significant trends  with mass, redshift or the morphological indicators were verified \cite{Applegate2016}. The $\gamma$ factor was taken to be $\gamma=0.848 \pm 0.085$ in agreement with the most recent estimates via observational data (SNe Ia, gas mass fraction and Hubble parameter) \cite{Rana2017} and in agreement with simulations \cite{Planelles2013}. We also use  priors on the $\Omega_{\rm b}$ and $\Omega_M$ parameters, i.e.,  $\Omega_{\rm b} = 0.049 \pm 0.0001$ and  $\Omega_{\rm M} = 0.315 \pm 0.007$, as given by current CMB experiments \cite{Aghanim2018}. These priors are from analyses by using exclusively CMB observations on the flat $\Lambda$CDM model.

	\section{Parametrizations}\label{sec:parameters}
	 The $\eta(z)$ functions considered here are  \cite{Holanda2010,Avgoustidis2010,Liang2013,Li2011}:
	
	\begin{equation}\label{eq:M1}
	\eta(z) = \eta_0, 
	\end{equation}
	
	\begin{equation}\label{eq:M2}
	\eta(z) = 1 + \eta_{1}z, 
	\end{equation}
	
	\begin{equation}\label{eq:M3}
	\eta(z) = 1 + \eta_{1}\frac{z}{1+z}, 
	\end{equation}
	
	\begin{equation}\label{eq:M4}
	\eta(z) = \eta_0 + \eta_{1}z, 
	\end{equation}
	
	\begin{equation}\label{eq:M5}
	\eta(z) = \eta_0 + \eta_{1}\frac{z}{1+z}.
	\end{equation}
	
	These are the main $\eta(z)$ functions widely used in the literature. Actually, they effectively parametrize our ignorance of the underlying process responsible for a possible CDDR violation. As commented on earlier, the current analyses cannot distinguish which $\eta(z)$ function better describes the data. Then, the basic idea here is to estimate the Bayesian evidence and compute the Bayes factor of the $\eta(z)$ functions with respect to $\eta=\eta_0$, which we verify to be $\approx 1$ within $2\sigma$ c.l. with the current SNe Ia and galaxy cluster data discussed in Sec. \ref{sec:data}. We will describe briefly what procedure follows to determine the Bayesian evidence and compare the $\eta(z)$ functions in the next section.
	
	\section{Bayesian Analysis}\label{sec:beyesian}
	
	Now, let us briefly introduce a summary on the Bayesian inference (BI). From the probability standpoint, BI is based on a measure of the degree of belief about a proposition. This method describes the connection between the competing models, the data, and the prior information concerning model parameters. The core of BI is the Bayes' theorem, which updates our preceding knowledge about the model in light of newly available data, being a consequence of the axioms of probability of theory. This hypothesis relates the posterior distribution $P(\Phi|D, M)$, likelihood $\mathcal{L}(D|\Phi, M)$,  the prior distribution $\pi(\Phi|M)$, and the Bayesian evidence $\mathcal{E}(D|M)$ \cite{Trotta2008}
	
	\begin{equation}\label{bayes}
	P(\Phi|D, M) = \frac{\mathcal{L}(D|\Phi, M) \pi(\Phi|M)}{\mathcal{E}(D|M)},
	\end{equation}
	where $\Phi$ is the set of parameters, $D$ represents the data and $M$ is the model.
	
	In the context of parameter constraint, the Bayesian evidence $\mathcal{E}(D|M)$ is just a normalization constant, and it does not affect the profile of posterior distribution since it does not depend upon the model parameters. However, it becomes an essential ingredient in the Bayesian model comparison viewpoint. So, the Bayesian evidence of a model in the continuous parameter space $\Omega$ can be written as 
	 
	\begin{equation}\label{evidence}
	\mathcal{E}(D|M) = \int_\Omega \mathcal{L}(D|\Phi, M) \pi(\Phi|M) d\Phi.
	\end{equation} 
	Therefore, the evidence is the average probability value across the allowed model parameter space before considering the data. 
	
	\begin{table}[H]
		\renewcommand{\arraystretch}{1.5}
		\renewcommand{\tabcolsep}{0.2cm}
		\centering
		\medskip
		\caption{\label{tab:jeffreys} The table shows the prior distribution of each parameter used in this work.}
		\begin{tabular}{|cc|} \hline
			$\ln{B_{ij}}$     & Interpretation \\ \hline
			Greater than $5$ 		        & Strong evidence for model i  \\
			$ [2.5,5] $         & Moderate evidence for model i       \\
			$ [1, 2.5]$         & Weak evidence for model i    \\
			$ [-1, 1]$          & Inconclusive       \\ 
			$ [-2.5, -1] $      & Weak evidence for standard model            \\
			$[-5, -2.5]$        & Moderate evidence for standard model             \\
			Less than $<-5$              & Strong evidence for standard model \\ \hline
		\end{tabular}
	\end{table}
	
	The most significant feature in the Bayesian model comparison is associated with the comparison of two models that describe the same data. The models fit the data well and are also predictive, shifting the average of the likelihood in Eq. (\ref{evidence}) in the direction of higher values. Instead, if a model which fits poorly or is not very predictive, the average of the likelihood decreases \cite{Liddle2007}. The application of Bayesian analysis has been widely applied in cosmology \cite{Hobson2002,Trotta2007,Santos2017,Lonappan2018,Kerscher2019,Silva2019,daSilva2019,Cid2019}. When comparing two models, $M_i$ versus $M_j$, given a set of data, we use the Bayes' factor defined in terms of the ratio of the evidence of models $M_i$ and $M_j$
	
	\begin{equation}\label{bayes_factor}
	B_{ij} = \frac{\mathcal{E}_i}{\mathcal{E}_j},
	\end{equation}
	where $\mathcal{E}_j$ is the standard model, and $\mathcal{E}_i$ are the competing models in which we want to compare.  
	Here, we will compare the $\eta(z)$ functions defined in the Sec. \ref{sec:parameters} by assuming that parametrization (\ref{eq:M1}) as the standard model (model $1$). The others parametrizations Eqs. (\ref{eq:M2}) , (\ref{eq:M3}), (\ref{eq:M4}) and, (\ref{eq:M5}), are named models $2$, $3$, $4$, and, $5$, respectively, in which we want to compare. If each model is assigned an equal prior probability, the Bayes factor gives the posterior odds of the two models.

	To quantify whether the model has favorable evidence or not, we adopted the Jeffreys scale showed in Table \ref{tab:jeffreys} to interpret the values of the Bayes factor in terms of the strength of the evidence in comparing two competing models. This scale was suggested by Ref. \cite{Trotta2008} as a revised and conservative version of the original Jeffrey scale \cite{Jeffreys61}. Note that this scale is empirically calibrated; i.e., it depends on the problem being investigated. Therefore, for an experiment for which $|\ln B_{ij}| < 1$, the evidence in favor of the model $M_i$ relative to model $M_j$ is interpreted as inconclusive. On the other hand, in the case of $\ln B_{ij} < -1$, we have support in favor of the model $M_j$. In this work, we consider the model $1$ as the reference model $M_j$. For a complete discussion about this scale, see Ref. \cite{Trotta2008}.
	
	Furthermore, we assume that both type Ia supernovae and galaxy clusters, and gas mass fraction datasets follow a Gaussian likelihood, such as 
	
	\begin{equation}\label{likelihood}
	\mathcal{L}(D|\Phi, M) \propto \exp\left[-\frac{\chi^2(D|\Phi, M)}{2}\right],
	\end{equation}
	where $\chi^2$ reads
	
	\begin{equation}
	\chi^2(D|\Phi, M) = \sum_i \left(\frac{\eta_{\text{obs}}(z_i) - \eta_{\text{mod}}(z_i)}{\eta^i_{\text{err}}}\right)^2,
	\end{equation}
	where $\eta_{\text{obs}}$ is a vector of the observed $\eta_{\rm obs}$ function defined by Eqs. (\ref{GasFrac5}) and (\ref{eq:DA-DL}); $\eta_{\text{mod}}$ are the theoretical values obtained from the parametrizations, Eqs. (\ref{eq:M1}), (\ref{eq:M2}), (\ref{eq:M3}), (\ref{eq:M4}), and, (\ref{eq:M5}) that we will test; and $\eta_{\text{err}}$ is the error given by propagation of uncertainty.
	
	In order to perform out the Bayesian analysis, we use \textsf{PyMultiNest} \cite{Buchner2014}, a Python module for \textsf{MultiNest} \cite{Feroz2008,Feroz2009,Feroz2013}, a generic tool that uses Importance Nested Sampling \cite{Skilling2004,Feroz2013} to calculate the evidence, but which still allows for posterior inference as a consequence. We plot and analyze the results using \textsf{GetDist} \cite{Lewis:2019xzd}. Additionally, to increase the efficiency in the estimate of the evidence, we chose to perform all analysis by working with a set of $2000$ live points,  so that the number of samples for posterior distributions was of the order $\mathcal{O}(10^4)$.
	
	It should be pointed out that BI depends on the priors distributions $ \pi(\Phi| M) $ adopted for the free parameters. This characteristic accounts for the predictive power of each model (parametrization), transforming this dependence in a property instead of a defect of the Bayesian inference framework. Albeit in the Bayesian analysis, the use of uniform (flat) priors can be acceptable in some cases; this type of prior can lead to issues of the point of view of model comparison. Uniform priors with distinct domain intervals change the evidence and can affect the Bayes factor between two competing models if it has not shared parameters. To use well-grounded priors, we considered values that reflect our actual state of knowledge about the parameters of the models investigated. Moreover, we assume the following flat priors on the set of parameters: $\eta_0 \sim $ Uniform$(0, 2)$ and $\eta_{1} \sim$ Uniform$(-1,1)$.

	\begin{table*}[t]
		\renewcommand{\arraystretch}{1.5}
		\renewcommand{\tabcolsep}{0.2cm}
		\centering
		\medskip
		\caption{\label{tab:results1}Confidence limits for the parameters using SNe Ia and galaxy clusters. The columns show the constraints on each model, whereas the rows show the parameter considering in this analysis.}
		\begin{tabular} {| c | c c c c c|}
			\hline
			Parameter &  Model 1 & Model 2 &Model 3 & Model 4 & Model 5 \\
			\hline
			{\boldmath$\eta_0           $} & $1.030\pm 0.017  $  & Fixed in 1  & Fixed in 1 & $1.030\pm 0.032    $ & $1.030\pm 0.037  $\\
			
			{\boldmath$\eta_1       	$} & - & $0.091\pm 0.059  $ & $0.134\pm 0.082   $ & $0.00\pm 0.11  $ & $0.00\pm 0.18   $\\
			
			{\boldmath$\ln \mathcal{E}  $} & $-17.992 \pm 0.041$ & $-17.134 \pm 0.032$ & $-16.696\pm 0.029$& $-19.910 \pm 0.048$ & $-19.465 \pm 0.047$ \\
			
			{\boldmath$\ln \mathcal{B}		$} & - & $0.858 \pm 0.052$  & $1.296\pm0.052$ & $-1.918\pm0.063$ & $-1.473 \pm 0.062$ \\
			{Interpretation} & - & Inconclusive & Weak evidence (favored)  & Weak evidence (against) & Weak evidence (against) \\
			\hline
		\end{tabular}
	\end{table*}

	\begin{table*}[t]
		\renewcommand{\arraystretch}{1.5}
		\renewcommand{\tabcolsep}{0.2cm}
		\centering
		\medskip
		\caption{\label{tab:results2} Confidence limits for the parameters using the gas mass fraction. The columns show the constraints on each model whereas the rows show the parameter considering in this analysis. Here, we marginalized $N$.}
		\begin{tabular} {| c | c c c c c|}
			\hline
			Parameter &  Model 1 & Model 2 &Model 3 & Model 4 & Model 5 \\
			\hline
			{\boldmath$\eta_0           $} & $0.977^{+0.025}_{-0.030}         $  & Fixed in 1 & Fixed in 1 & $0.980^{+0.025}_{-0.031} $ &$ 0.982^{+0.025}_{-0.031}            $\\
			
			{\boldmath$\eta_1       	$} & - & $-0.008\pm 0.020  $ & $-0.025\pm 0.033             $& $-0.006\pm 0.019            $ & $-0.020\pm 0.034  $\\
			
			{\boldmath$\ln \mathcal{E}  $} & $-38.540 \pm 0.049$ &  $-39.233 \pm 0.050$  & $-38.521 \pm 0.047$& $-42.256 \pm 0.063$ & $-41.560 \pm 0.061$ \\
			
			{\boldmath$\ln \mathcal{B}	$} & - & $-0.693\pm0.070$ & $0.019\pm0.068$ & $-3.716\pm 0.080$ & $-3.020\pm0.078$ \\
			{Interpretation} & - & Inconclusive & Inconclusive &  Moderate evidence (against)  & Moderate evidence (against) \\
			\hline
		\end{tabular}
	\end{table*}

	\section{Results}\label{sec:results}

	The results achieved considering the SNe Ia and $D_{\rm A}$ from galaxy clusters data are shown in Fig. \ref{fig:regions1}. As shown in the Fig. \ref{fig:regions1}(a), the vertical traced line means $\eta_0 = 1$, i.e., CDDR validity. By considering the data, we obtain $\eta_0 = 1.030\pm 0.017$ for model 1. This value obtained is compatible in $2\sigma$ C.L. with $\eta_0 = 1$ (light green region). In Fig. \ref{fig:regions1}(b), we show the results for models 2 and 3. Now, the vertical traced line means $\eta_1 = 0$ (CDDR validity). The values obtained for $\eta_1$ were $\eta_1 = 0.091\pm 0.059$ for model 2 and $\eta_{1} = 0.134\pm 0.082$ for model 3. See that only model 2 is compatible with $\eta_1 = 0$ in $2\sigma$ of confidence (light green region), and the model 3 is compatible in $3\sigma$ with $\eta_1 = 0$ (light blue region). Finally, in Fig. \ref{fig:regions1}(c), we present the triangle plot composed of the regions of confidence for $\eta_0$ and $\eta_{1}$ and the posteriors distributions for models 4 and 5. The traced lines mean the values in which the CDDR is valid ($\eta_0=1$ and $\eta_1 = 0$). The values obtained for the parameters of the model 4 were  $\eta_0 = 1.030\pm 0.032$ and $\eta_1 = 0.00\pm 0.11$. These values are compatible in $2\sigma$ C.L. with the validity of CDDR. For model 5, we obtained  $\eta_0 = 1.030\pm 0.037$ and $\eta_1 = 0.00\pm 0.18$; they are also compatible in $2\sigma$ C.L. with the validity of CDDR. We can see there is an anticorrelation between the parameters.  Note that the data considered constrain the parameters of model 4 better than model 5.

	In Fig. \ref{fig:regions2}, we show the results obtained considering the x-ray gas mass fraction of galaxy clusters and SNe Ia. The model 1 is consistent in $1\sigma$ confidence with CDDR validity, $\eta_0 = 0.977^{+0.025}_{-0.030}$. In the case of  models 2 and 3, Fig. \ref{fig:regions2}(b), we obtained that ones are consistent in $1\sigma$ confidence with $\eta_1 = 0$. In the Fig. \ref{fig:regions2}(c), we show the corner plot for models 4 and 5. The horizontal gray line means $\eta_{1} = 0$, and the vertical line is $\eta_0 = 1$. The values obtained for the parameters of model 4 were  $\eta_0 = 0.980^{+0.025}_{-0.031}$ and $\eta_1 = -0.006\pm 0.019$ and model 5 were $\eta_0 =0.982^{+0.025}_{-0.031}$ and $\eta_1 = -0.020\pm 0.034  $. These values are compatible with CDDR validity in $1\sigma$ C.L.

	For the sake of Bayesian model comparison, we estimate the values of the logarithm of the Bayesian evidence ($\ln \mathcal{E}$) and, the Bayes factor ($\ln \mathcal{B}$), Tables \ref{tab:results1} and \ref{tab:results2}. These results were obtained considering the priors defined in last section, and we considered model $1$ as the reference one. In the case of $D_{\rm A}$ from galaxy clusters and SNe Ia data, Table \ref{tab:results1}, we first observe that model $2$ has a positive value of Bayes factor ($\ln \mathcal{B} = 0.858 \pm 0.052$). According to the Jeffreys scale, Table \ref{tab:jeffreys}, we can conclude that this model has evidence inconclusive concerning to model 1. By considering model $3$, we obtain $\ln \mathcal{B} = 1.296 \pm 0.052$, so this model was
    weakly supported by the data. Regarding models $4$ and $5$, we obtain negative values for the Bayes factor, by which we mean that they have weak evidence unsupported by the
    data. Thus, we conclude that model $3$ is weakly favored by $D_{\rm A}$ from galaxy clusters and SNe Ia data.
	
	By considering the second dataset, i.e., gas mass fraction and SNe Ia, we also implement Bayesian model comparison, Table \ref{tab:results2}. Model $1$ is the reference one. Models $2$ and $3$ have evidence inconclusive regarding the data. Concerning the other models, we note that they have moderate evidence disfavored by the data. From the Bayesian comparison model analysis point of view and the data considered, we conclude that all models have inconclusive and moderate evidence disfavored by x-ray gas mass fraction.
	
	In Ref. \cite{Hees2014}, it was shown that for theories motivated by scalar-tensor theories of gravity, which introduce an additional coupling between the Lagrangian of the usual nongravitational matter field with a new scalar field, variation in the value of the fine structure constant,  modification of the CDDR and modifications of the CMB temperature evolution law  are intimately and unequivocally linked. In this way, other possibilities to probe some departure from the CDDR validity is by  using the CMB  spectral distortion, which can be constrained using Cosmic Background Explorer (COBE)/Far-Infrared Absolute Spectrophotometer (FIRAS) or the rescaling of the CMB temperature law (see Refs. \cite{Battistelli2002,Luzzi2009,Chluba2014} for more details).
	
	\section{Conclusions}\label{sec:conclusions}
	
	In the last ten years, several works tested the cosmic distance duality relation , $\frac{D_{\rm L}(z)}{D_{\rm A}(z)(1+z)^{2}} =1$ , by considering a deformed CDDR, such as $\frac{D_{\rm L}(z)}{D_{\rm A}(z)(1+z)^{2}} = \eta(z)$. Several $\eta(z)$ functions were considered, however,  the current analyses could not distinguish which $\eta(z)$ function better describes the data. 
	
\begin{figure*}[t]
		\centering
		\includegraphics[scale=0.30]{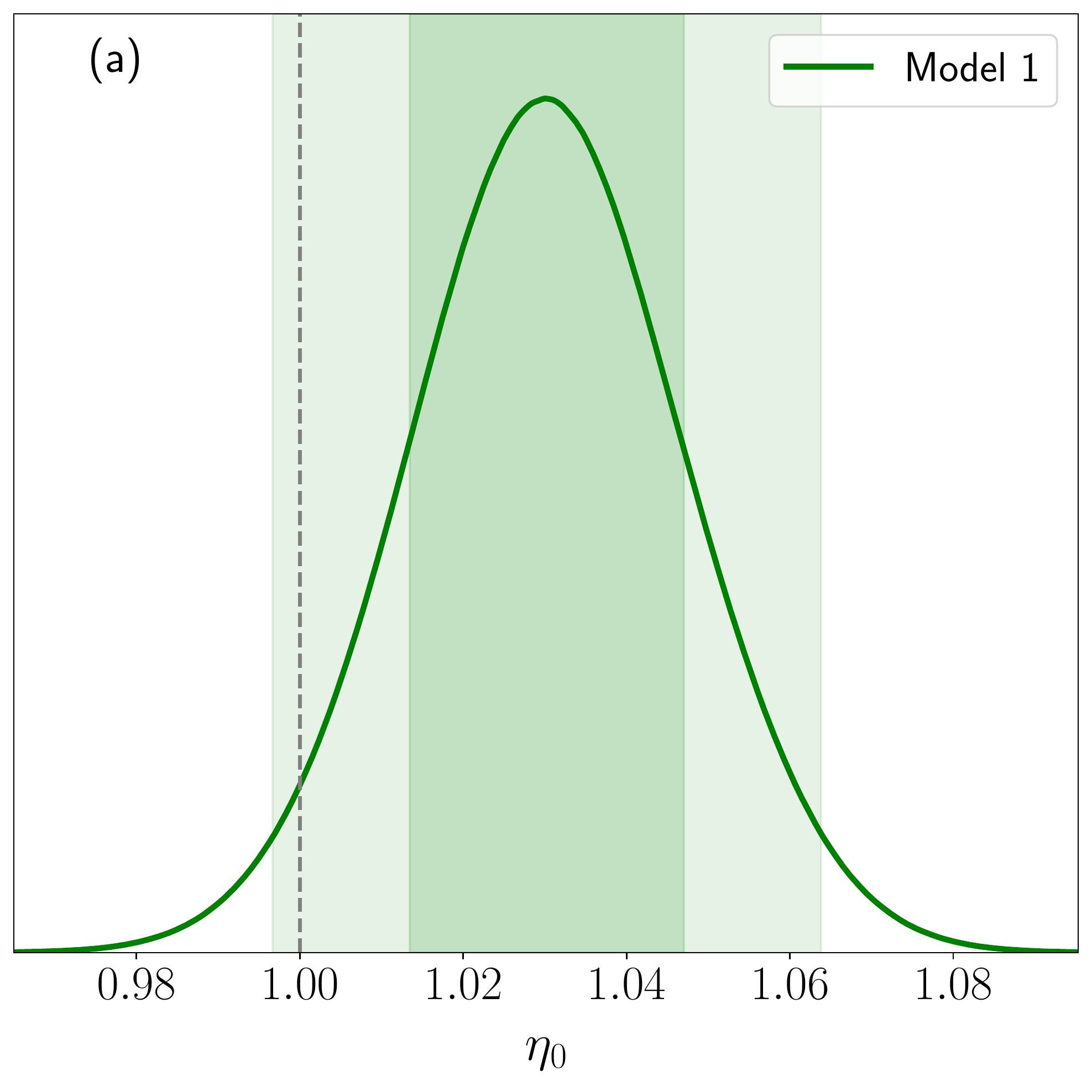}
		\includegraphics[scale=0.30]{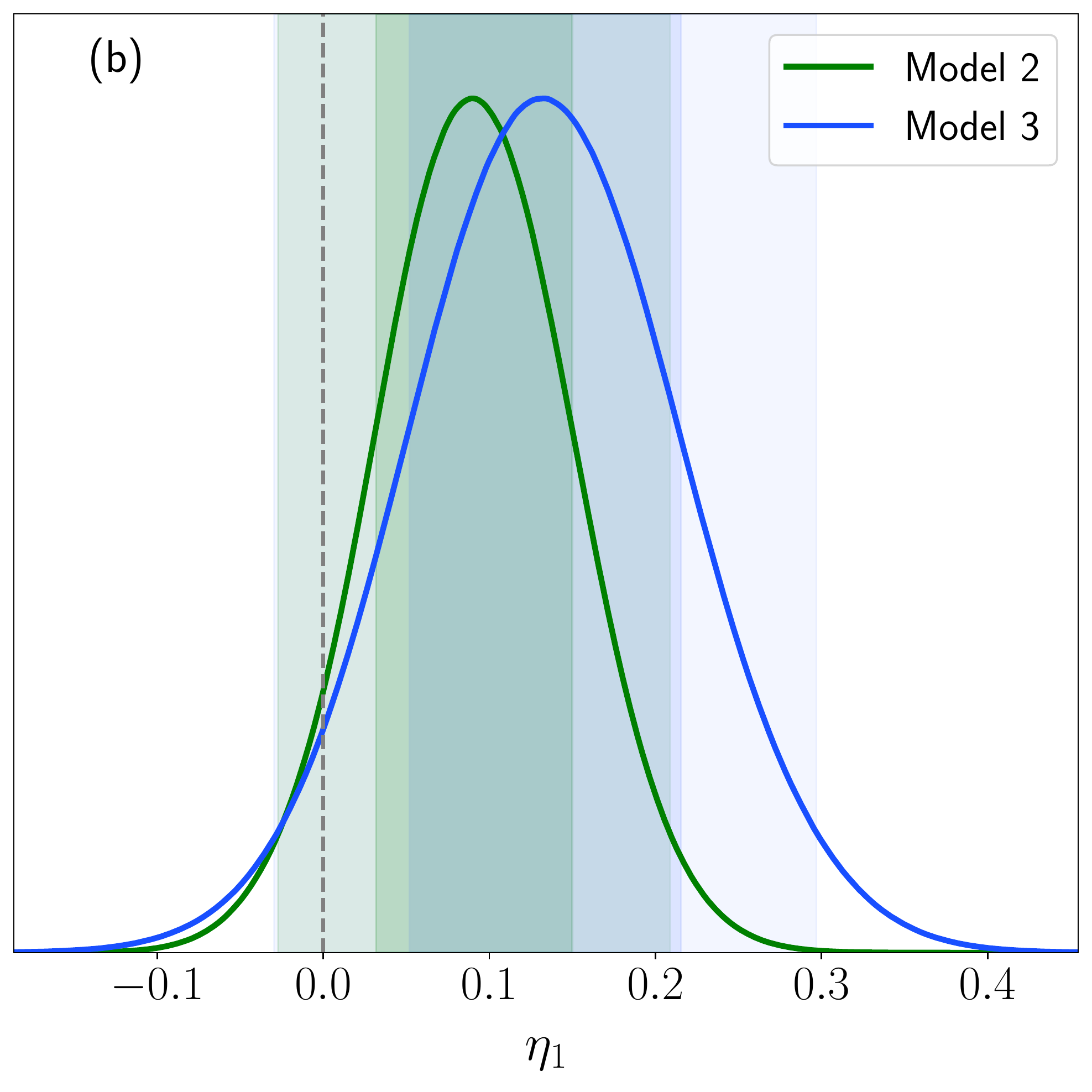}
		\includegraphics[scale=0.40]{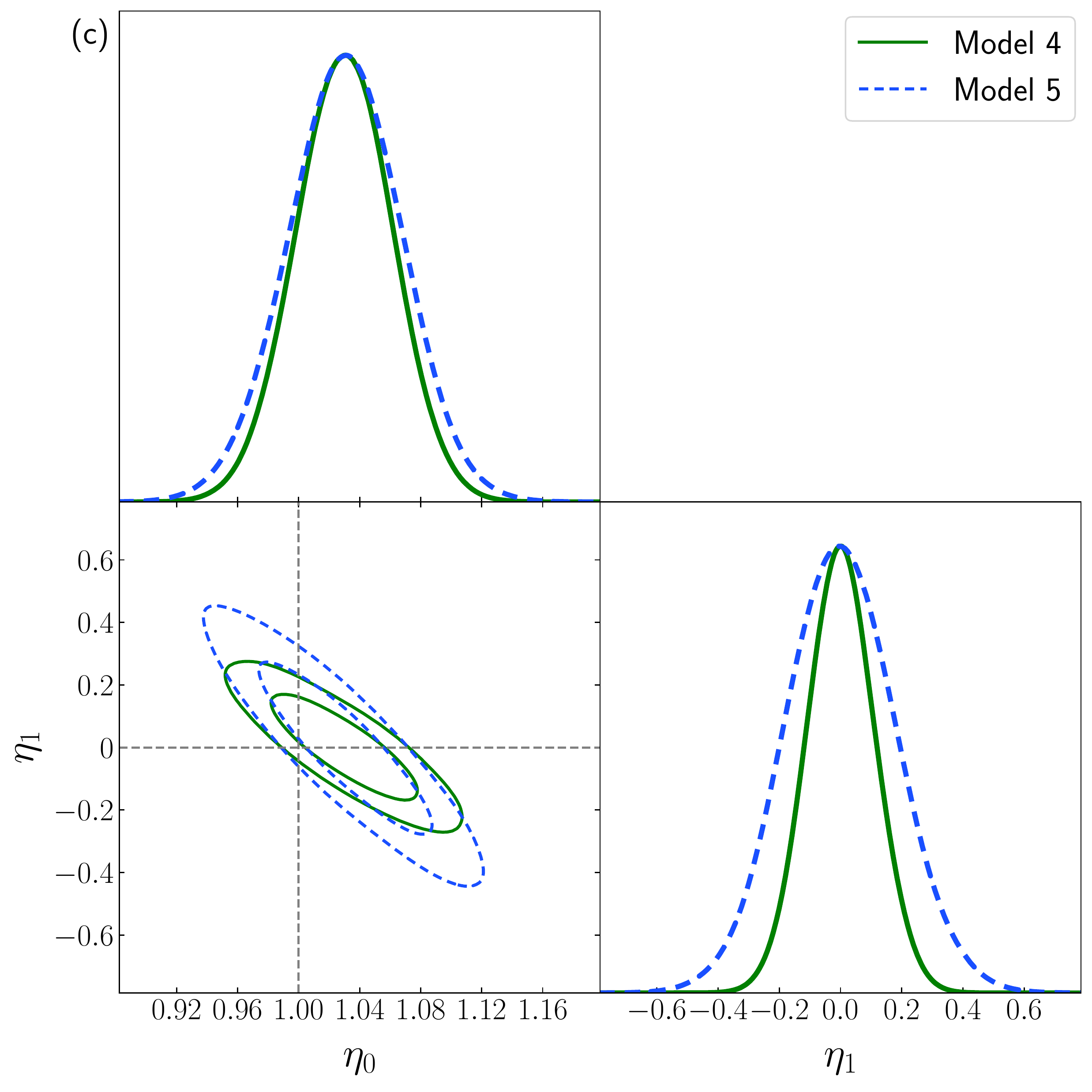}
		\caption{\label{fig:regions1} The posteriors distributions for SNe Ia and galaxy cluster data. The first row shows the reference model (part (a)) and, the models 2 and 3 (part (b)). The last row shows the corner plot for models 4 and 5 (part (c)).}
	\end{figure*}

\begin{figure*}[t]
		\centering
		\includegraphics[scale=0.30]{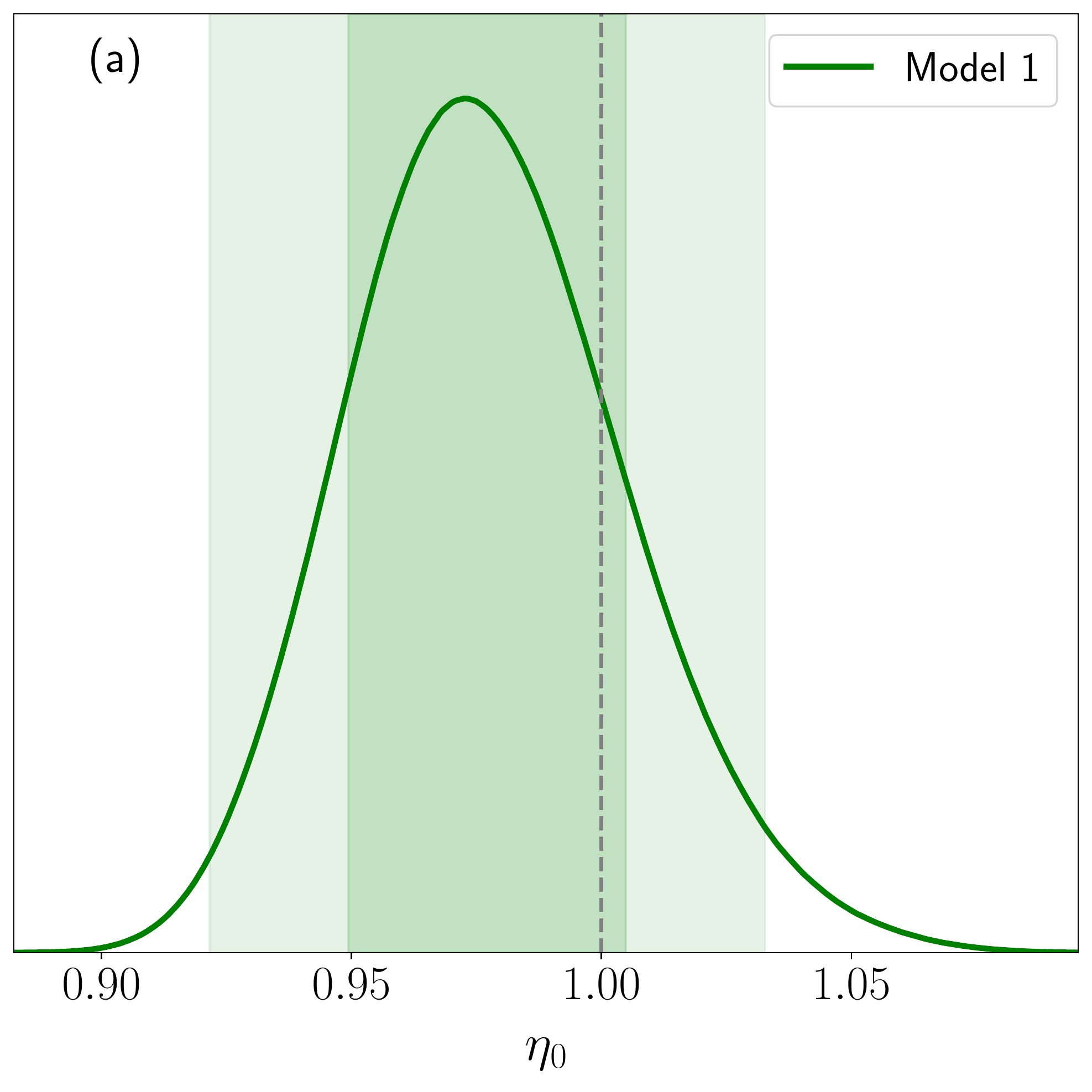}
		\includegraphics[scale=0.30]{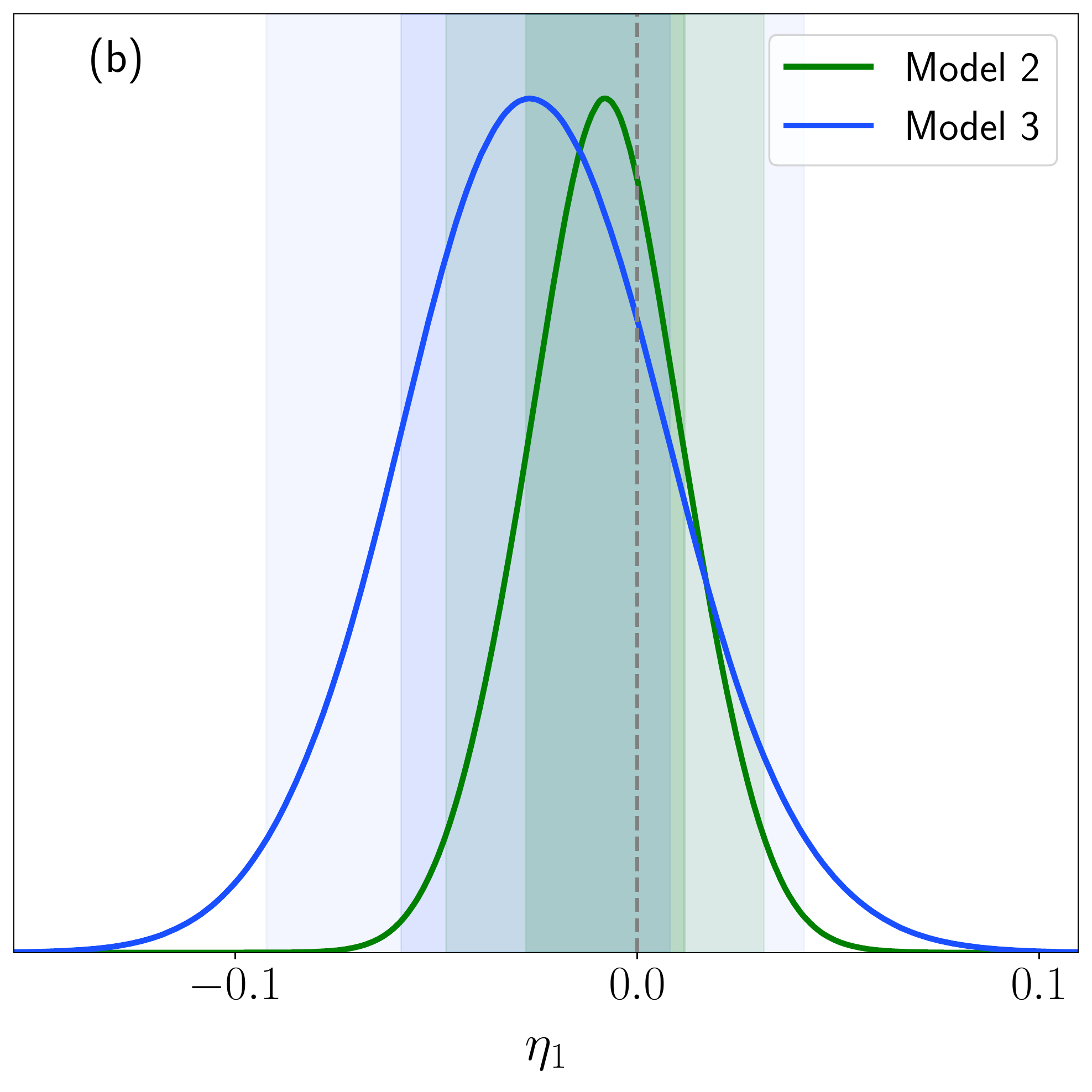}
		\includegraphics[scale=0.40]{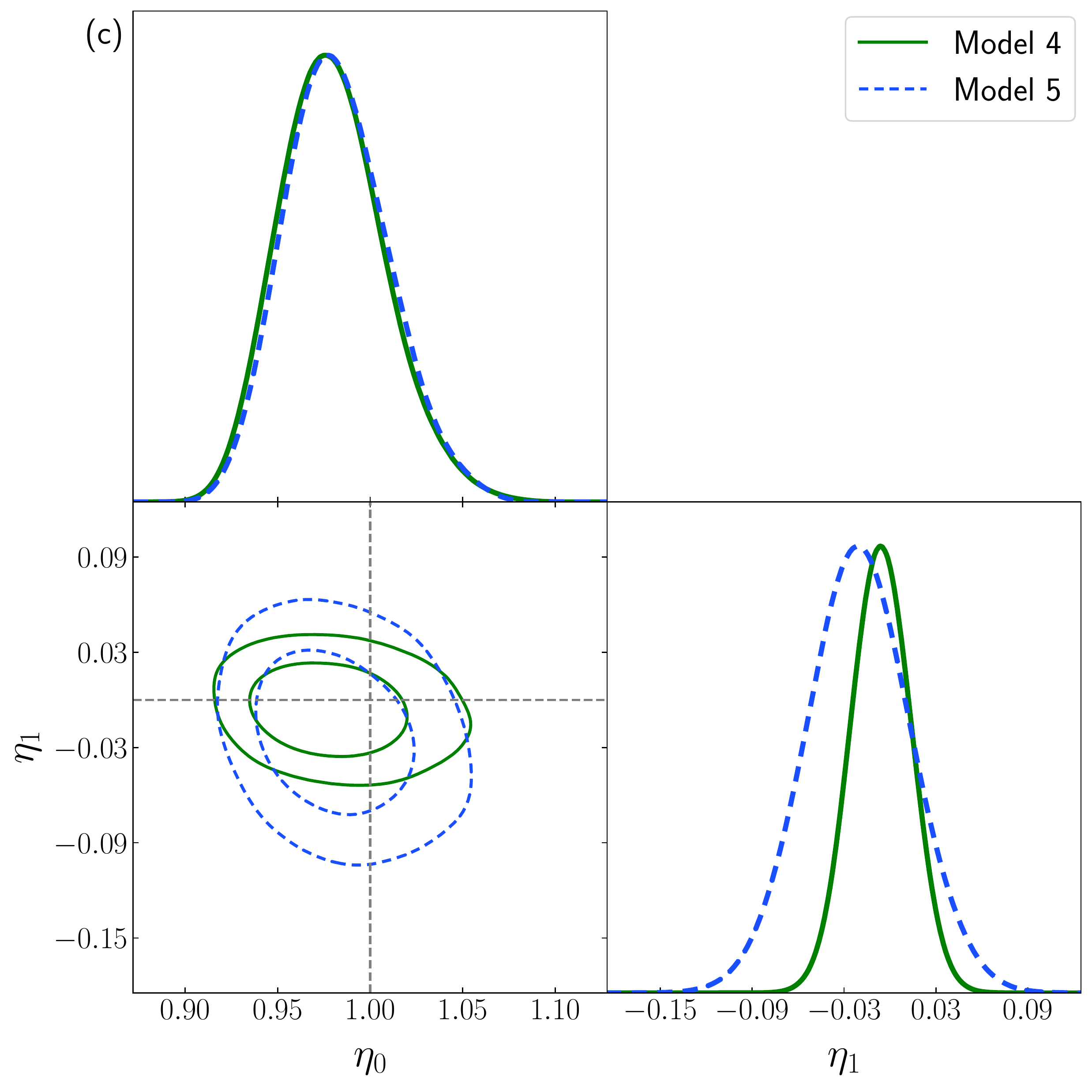}
		\caption{\label{fig:regions2} The posteriors distributions for gas mass fraction. The first row shows the reference model (part (a)) and, the models 2 and 3 (part (b)). The last row shows the corner plot for models 4 and 5 (part (c)).}
	\end{figure*}

	In this work, we relaxed the CDDR by assuming the $\eta(z)$ functions as given in Sec. \ref{sec:parameters}. In order to decide which $\eta(z)$ function better  describes the data, we implemented a Bayesian inference analysis in terms of the strength of the evidence according to Jeffreys scale, Table \ref{tab:jeffreys}. We considered the priors defined in the Sec. \ref{sec:beyesian} and astronomical data such as SNe Ia, diameter distance angular of the galaxy clusters and, x-ray gas mass fraction. The results obtained are reported in Tables \ref{tab:results1} and \ref{tab:results2}, where we showed the mean, the error, the Bayesian evidence and, the Bayes factor for all models studied here. In Figs. \ref{fig:regions1} and \ref{fig:regions2}, we showed the $1\sigma$ and $2\sigma$ regions of confidence and the posteriors distributions for all models. 
	
    The statistical constraints on all the functions  implied  that the CDDR remains valid in $1\sigma$ in the analyses by using SNe Ia and galaxy cluster gas mass fractions and in $2\sigma$ C.L. when $D_{\rm A}$ from galaxy clusters and SNe Ia data were considered. However, we concluded from the Bayesian comparison that  $\eta(z)=1+\eta_0z/(1+z)$ was weakly favored in the CDDR test considering the $D_{\rm A}$ from galaxy clusters and SNe Ia data with respect to our standard model $\eta(z)=\eta_0$. On the other hand, in the CDDR test considering the galaxy cluster gas mass fractions and SNe Ia, all the $\eta(z)$ functions had inconclusive evidence or moderate evidence(against) with respect to our standard model. In both 
    methodologies, $\eta(z)=\eta_0=1$ is in agreement within $2\sigma$ C.L. with the data. 
	
	Finally, we concluded that the present data used in our analyses failed to provide which function of $\eta(z)$ better describes the evolution of the CDDR with redshift. Probably, this is a consequence of the galaxy cluster dataset used in this paper, and they still have large statistical and systematic errors ($\approx 20\%$). We believe that when applied to upcoming galaxy cluster data, the analyses proposed here may be useful to probe a possible violation of the CDDR.

	\begin{acknowledgements}
		The authors thank CAPES and CNPq, Brazilian scientific support federal agencies, for financial support, and High Performance Computing Center (NPAD) at UFRN for providing the computational facilities to run the simulations. W. J. C. da Silva acknowledges financial support from the Programa de Capacitação Institucional PCI/ON/MCTI. R. F. L. Holanda thanks financial support from Conselho Nacional de Desenvolvimento Cientıfico e Tecnologico (CNPq) (No. 428755/2018-6 and 305930/2017-6).
	\end{acknowledgements}

	\bibliographystyle{apsrev4-1}
	\bibliography{references}

\end{document}